\documentclass[preprint]{aastex}

\newcommand\OUCHII{UCH\,{\sc ii}}
\newcommand\UCHII{UCH}
\newcommand\HII{H\,{\sc ii}}

\newcommand\kms{km~s$^{-1}$}

\newcommand\cmthree{cm$^{-3}~$}

\newcommand\dycm{dyne cm$^{-2}~$}
\newcommand\etal{et al.}
\newcommand\be{\begin{equation}}
\newcommand\ee{\end{equation}}
\newcommand\bea{\begin{eqnarray}}
\newcommand\eea{\end{eqnarray}}

\catcode`\@=11
\newcommand{\gsim}{${\mathrel{\mathpalette\@versim>}}$}
\newcommand{\lsim}{${\mathrel{\mathpalette\@versim<}}$}
\newcommand{\@versim}[2]{\lower 2.9truept \vbox{\baselineskip 0pt \lineskip
    0.5truept \ialign{$\m@th#1\hfil##\hfil$\crcr#2\crcr\sim\crcr}}}
\catcode`\@=12

\newcommand\bbeta{\boldmath$\beta$}

\shorttitle{Carbon recombination lines toward \OUCHII\ regions}
\shortauthors{D. Anish Roshi et al.}

\begin{document}

\title{An 8.5 GHz Arecibo survey of Carbon Recombination Lines  
toward Ultra-compact \HII\ regions: Physical properties of dense molecular material}

\author{D. Anish Roshi}
\affil{Raman Research Institute, Bangalore, India 560 080 and \\
National Radio Astronomy Observatory\altaffilmark{1}, Green Bank, WV 24944, USA}
\email{anish@rri.res.in}

\author{Dana. S. Balser}
\affil{National Radio Astronomy Observatory, Green Bank, WV 24944, USA }
\email{dbalser@nrao.edu}

\author{T. M. Bania}
\affil{Institute for Astrophysical Research, 725 Commonwealth Avenue, \\
             Boston University, Boston, MA 02215, USA.}
\email{bania@bu.edu}

\author{W. M. Goss}
\affil{National Radio Astronomy Observatory, Socorro,  NM 87801, USA }
\email{mgoss@nrao.edu}

\and

\author{C. G. De Pree}
\affil{Department of Physics and Astronomy, Agnes Scott College, 
141 East College Avenue, Decatur, GA 30030. }
\email{cdepree@agnesscott.edu}

\altaffiltext{1}{The National Radio Astronomy Observatory is a 
             facility of the National Science Foundation, operated 
             under a cooperative agreement by Associated 
             Universities, Inc.}

\begin{abstract}
We report here on a survey of carbon recombination lines (RLs) near 8.5 GHz toward
17 ultra-compact \HII\ regions (\UCHII s). Carbon RLs are detected in 11 directions,
indicating the presence of dense photodissociation regions (PDRs) associated
with the \UCHII s. In this paper, we show that the carbon RLs provide
important, complementary information on the kinematics and physical properties
of the ambient medium near \UCHII s. Non-LTE models for 
the carbon line forming region are developed, assuming that the PDRs 
surround the \UCHII s, and we constrained the model parameters   
by multi-frequency RL data. Modeling shows that 
carbon RL emission near 8.5 GHz is dominated by stimulated emission and
hence we preferentially observe the PDR material that is in front of the \UCHII\
continuum. We find that the relative motion between ionized gas and the
associated PDR is about half that estimated earlier, and has
an RMS velocity difference of 3.3 \kms.
%+/- 2.9 km/sec.  
Our models also give estimates for the PDR density and pressure.  
We found that the neutral density of PDRs is typically 
$>$ 5 $\times$ 10$^5$ \cmthree 
and \UCHII s can be embedded in regions with high ambient pressure. 
Our results are consistent with a pressure confined \HII\ region model
where the stars are moving relative to the cloud core.  Other models
cannot be ruled out, however.  Interestingly,
in most cases, the PDR pressure is an order of magnitude larger
than the pressure of the ionized gas. Further investigation is
needed to understand this large pressure difference.
\end{abstract}

\keywords{
ISM:\HII\ regions --- ISM: general --- line:formation --- radio lines:ISM --- 
stars:formation --- surveys  
 }

\section{Introduction}

Observations of molecular lines toward \UCHII s
have shown that the natal clouds harboring the \HII\ regions
have densities $>$ 10$^5$ \cmthree\ and temperatures
between 100 -- 200 K (see \nocite{c02}Churchwell 2002).
The presence of such dense clouds has other 
observable effects. In particular, far-ultraviolet (FUV) photons 
(6.0 -- 13.6 eV) from OB stars should produce photo-dissociation regions 
(PDRs) in the neutral material close to the \UCHII\
(see review by \nocite{ht97}Hollenbach \& Tielens 1997). 
Gas phase carbon in these regions is ionized by photons
in the energy range 11.3 -- 13.6 eV.  
The physical conditions in these PDRs are ideally suited for 
producing observable radio recombination lines (RLs)
of carbon (\nocite{nwt94}Natta, Walmsley \& Tielens 1994; 
\nocite{jetal05}Jeyakumar \etal\ 2005). 

To date, most objects that have been searched for carbon
RL emission are extended \HII\ regions. These \HII\ regions are 
often density bound and the molecular densities near them are low. 
Thus, not surprisingly, these searches have detected carbon lines 
only toward a few extended \HII\ regions. \UCHII s  form in or near 
dense molecular material and hence they appear to be ideal targets 
for detecting carbon RLs.  A handful of recent VLA 
observations was successful in detecting 
carbon RLs toward \UCHII s. For example, observations
toward W48A and W49 have detected, respectively, the 76$\alpha$ (14697.314 MHz) 
and 92$\alpha$ (8313.528 MHz) transitions of carbon
(\nocite{rgaj05}Roshi \etal\ 2005, 
\nocite{detal05}De Pree \etal\ 2005). 
The angular resolutions of these observations
(5\arcsec ~and 2\arcsec ~respectively) were sufficient to infer that
the line emission is associated with the \UCHII s.  
Moreover, the detected RLs have velocities similar to that of
the molecular clouds harboring the \UCHII s, implying that 
the line forming region is embedded in the dense material 
near the \HII\ regions.

Multi-frequency carbon RL observations from PDRs associated with 
\UCHII s can be used: (1) to estimate the physical 
properties of the dense molecular material 
(\nocite{jetal05}Jeyakumar \etal\ 2005) and
(2) to test whether \UCHII s are pressure 
confined (\nocite{rgaj05}Roshi \etal\ 2005).
With these aims, we made a survey of carbon RLs near 8.5 GHz with the Arecibo 
telescope\footnote{Arecibo Observatory is part of the National Astronomy 
and Ionosphere Center (NAIC), which is operated by Cornell University 
under a cooperative agreement with the National Science Foundation.} 
toward 17 \UCHII s. The source selection was made such that at least one source
from each of the six morphological types suggested by 
\nocite{wc89b}Wood \& Churchwell (1989b) and 
\nocite{kcw94}Kurtz, Churchwell \& Wood (1994) for \UCHII s 
is observed in the survey
(see Table~\ref{tab1}). Another criterion used for selection was that  
the continuum flux densities at 2 cm of the \UCHII s are $>$ 150 mJy
(\nocite{wc89b}Wood \& Churchwell 1989b; \nocite{kcw94}Kurtz \etal\ 1994).
%The details of the observations, 
%data analysis and modeling of the line forming region 
%are presented in this paper. 
In \S\ref{sec:obs} we discuss the observations and data analysis procedure. 
The results of the survey are given in \S\ref{sec:result}.  
In \S\ref{sec:model} we present the details of modeling of the carbon RL
emission toward a selected subset of \UCHII s from our sample. 
Our conclusions are given in \S\ref{sec:con}.

\section{Observations and Data Analysis}
\label{sec:obs}

The observations were made at frequencies near 8.5 GHz with the Arecibo Telescope 
on May 30 through June 4 2003. At the observed frequency, the telescope
has a beam-width (FWHM) of 28\arcsec. The wide bandwidth of the 8 to 10 GHz 
receiver system of the telescope was used to simultaneously observe
four RL transitions (89$\alpha$, 90$\alpha$, 91$\alpha$ and 92$\alpha$)
of carbon, hydrogen and helium from two polarizations. In addition to
these lines, the H113$\beta$ line (8878.730 MHz) was present in the
frequency band of the spectrometer used for observing the 90$\alpha$ RL
transitions. The auto-correlation spectrometer was configured in the four sub-band mode
to observe the four RL transitions. The transitions were selected such that the
frequency range was free of radio frequency interference (RFI). 
%Moreover, these transitions are located in the frequency range  
%where the telescope gain is maximum. 
% Arecibo's maximum gain is at 430 MHz.....
We used 25 MHz bandwidth for each sub-band, which corresponds to a velocity 
range of $\sim$ 840 \kms. This velocity range was adequate
to observe RLs of carbon, helium and hydrogen in each sub-band. We selected 9 bit 
sampling in the correlator to minimize effects due to RFI and to maximize the 
sensitivity of the observations. In the chosen configuration, the correlator provides
1024 spectral channels for each sub-band and for each polarization.
Before observing the target sources, the setup was tested by observing
the `extended' \HII\ region G43.17+0.0 (\nocite{l89}Lockman 1989).

Our observations were one of the first scientific
uses of the 8 -- 10 GHz system and hence there were several factors
affecting telescope performance. Some of these factors are: (1)
an inaccurate telescope pointing model and (2) irregularities of the 
primary surface panel settings. We, moreover, 
had to use a `gain curve' which is poorly sampled in azimuth and
elevation. The gain curve essentially provides the telescope gain
(i.e. antenna temperature in K per Jy) for different azimuths 
and zenith angles. This curve is needed to calibrate the
observed spectra in Jy, since the gain of the Arecibo telescope 
is a function of azimuth and zenith angle. The gain curve is 
derived from observations toward a set of calibrator sources. 
We had to use data toward calibrator sources observed during
the period 01 March to 19 July, 2003 to derive the gain curve. 
The number of calibrators observed in this 
period was not large enough to provide good sampling of the 
telescope gain over azimuth and zenith angle. 
(Unfortunately the receiver system was modified a few weeks after our 
observations so we could not improve upon this gain curve.)
During the 15 months subsequent to our observations the 
performance of the 8--10 GHz system, the data quality
and calibration procedure have all been significantly improved. 
Due to the performance limitations during our observations, however, 
we were forced to devise several analysis procedures in order
to minimize the calibration errors. In addition to the software provided
by Arecibo observatory, we developed programs in IDL to implement
these procedures.

Since the telescope pointing model was not accurate we observed  
calibrator sources B1843+098, B1857+129 and B2018+295 at intervals 
of about 45 minutes.  We made azimuth--zenith angle continuum cross
scans in order to measure pointing offsets.  These offsets were used 
to correct the telescope pointing before observing the target sources. 
We found that the pointing offset was as large as 16\arcsec\ 
($\sim$ 0.6 $\times$ FWHM beam width) in some cases.  Because the position could,
of course, not be corrected during observations pointing errors have 
resulted in an additional uncertainty in the amplitude calibration of our spectra.  
We estimate below that the RL amplitudes reported here have an error of 20\%.  

Careful bandpass calibration is important for 
our observations since the RL flux density is   
less than a percent of the system equivalent flux density. 
To do this we observed a
target source for 5 minutes, then switched 
to an off-source position and acquired data for the 
same amount of time.  The off-source position was selected 
such that it followed the same azimuth--zenith angle track
as that of the target source during the integration. 
After each on- and off-source measurement pair, 
spectra were taken with a noise calibration signal, turned on 
and off for 30 sec, respectively. These data were used 
to derive the system temperature. 

We configured the spectrometer to have an integration time of 1 second. 
This short averaging time was selected in case the spectra
were affected by RFI. Examination of the data showed that the four 
bands were not affected by any RFI. The spectrometer provides the 
Fourier transforms of the measured auto-correlation function after 
applying uniform weighting. The spectral resolution achieved is about 
1 \kms\ which is adequate to resolve the carbon RLs.  

The noise calibration scans and the gain curve were used to calibrate the 
spectrum in units of Jy. We noticed that the total power in the 1 second
integrations was not stable for some of the sub-bands.  Therefore we visually 
examined the calibrated spectra for relatively high RMS and discrepant spectral 
baselines and excised these data.  Less than 1\%\ of the data were excised.

\begin{figure}
\plotone{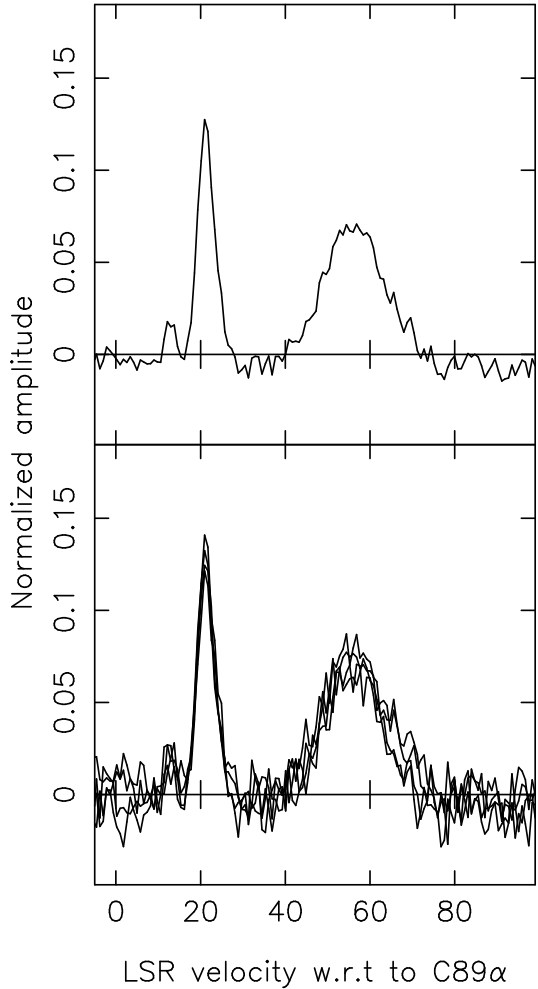}
\caption{A demonstration of the spectral resampling procedure.
The final average spectrum obtained toward G61.48+0.09 
is shown on the top. The four spectra plotted in the bottom panel correspond
to the RL transitions 89$\alpha$, 90$\alpha$, 91$\alpha$ and 92$\alpha$. They
are obtained by averaging the data for each sub-band separately and by resampling
at the same velocity resolution as that of the spectrum shown on top.  
The velocity range is restricted to show the carbon RL (narrow line feature), 
which has the smallest line width (4.7 \kms) and hence is affected most 
by any error in the resampling process. The broad line feature is the
helium RL. The good agreement between the RL profiles shows the efficacy of the
resampling process.
\label{fig0} }
\end{figure}

The velocity resolution of the sub-bands were not the same because 
they are centered at different sky frequencies. For the carbon RLs 
observed here the velocity resolution can change by as much as 10\%. 
Moreover the spectra of the four sub-bands are not sampled at the 
same velocities.  Since the spectra were obtained by Fourier transforming
the measured autocorrelation functions with uniform weighting, the filter 
response of the spectrum was a sinc function. We therefore resampled the 
spectra using sinc interpolation in order to have an identical velocity 
sampling for all sub-bands. We referenced all sub-bands to the velocity
scale and resolution of the spectrum corresponding to the 89$\alpha$ 
transition. A demonstration of our resampling process is 
given in Fig.~\ref{fig0}, where the average spectrum obtained toward G61.48+0.09
and the resampled average spectra of the four sub-bands are shown. 
The velocity range is restricted to show the carbon RL, which has the smallest
line width (4.7 \kms) and hence is affected most by any error in the resampling 
processes.  The good agreement between the carbon RL profiles proves the 
efficacy of the resampling process. The four resampled spectra of a scan 
were then averaged to produce a single calibrated spectrum 
for each polarization.

We examined the amplitude of the hydrogen RLs in the individual on/off
pairs for each source. For some
sources the hydrogen RL amplitude changed during the integration.
Since the RL intensity is not expected to change, this variation in line 
amplitude is due to pointing errors and/or errors in the estimated antenna 
gain.  To minimize the error in the final spectrum and to maximize the 
signal-to-noise ratio, we weighted the spectrum of each scan by the ratio of 
the hydrogen RL amplitude to the spectral variance before averaging. 
The variance was estimated using spectral values in the velocity range
where RLs were not present and after removing a residual baseline.
During the observations we choose to apply a Doppler correction for each sub-band 
separately and hence the spectra corresponding to different scans could be
averaged without any further velocity shift. The frequency offset used for each
sub-band was based on the rest frequency of the carbon RL observed in that sub-band. 

The spectral baselines of the final calibrated spectra were fit with
a third-order (or less) polynomial model using only the line-free
channels.  The continuum flux density was determined to be the mean
value of this baseline model.  The spectral line profiles were modeled
using a minimum number of Gaussian components that were able to fit
the hydrogen, helium, carbon, and H133$\beta$ lines without any
constraints on the model parameters.  In one object there was evidence
for a weak heavy ion (heavier than C$^+$ ion) RL in the residuals.  
The Gaussian parameters of
this profile were obtained by keeping the model parameters of the
other four RLs constant.

The error in the line flux density in the final average spectrum can be 
estimated by comparing the hydrogen RL amplitude in the final spectrum 
with that obtained from individual scans. The comparison showed that 
the line flux density was within 15--18\% of the maximum RL amplitude 
observed for a source but clearly biased toward lower
values. However, the spectrum with maximum amplitude itself may not 
have the correct flux density in all cases, since sampling of the 
gain curve is non-optimal. So we further compared the continuum
flux densities of two sources (G32.80+0.19 and G70.29+1.60) with those
obtained from existing 3.6 cm interferometric images (see Table~\ref{tab3}). 
The continuum flux densities obtained from our observations (see Table~\ref{tab2}) 
were measured with respect to 
the off-source position used for bandpass calibration. The two sources 
were selected based on negligible sky contribution to the continuum emission 
at the off-source position, as inferred from existing continuum surveys 
(\nocite{retal90}Reich \etal\ 1990; \nocite{aetal79}Altenhoff \etal\ 1979). 
Neglecting the small wavelength difference, the flux densities obtained
from our observations are within 10\% of those estimated from interferometric
data. Based on this comparison and the lower line flux densities in the
final spectra compared to the observed maximum values, we estimate a 
$1\sigma$ error of 20\% for all RL flux densities obtained from our survey data.

\section{Results}
\label{sec:result}

\begin{figure}
\plotone{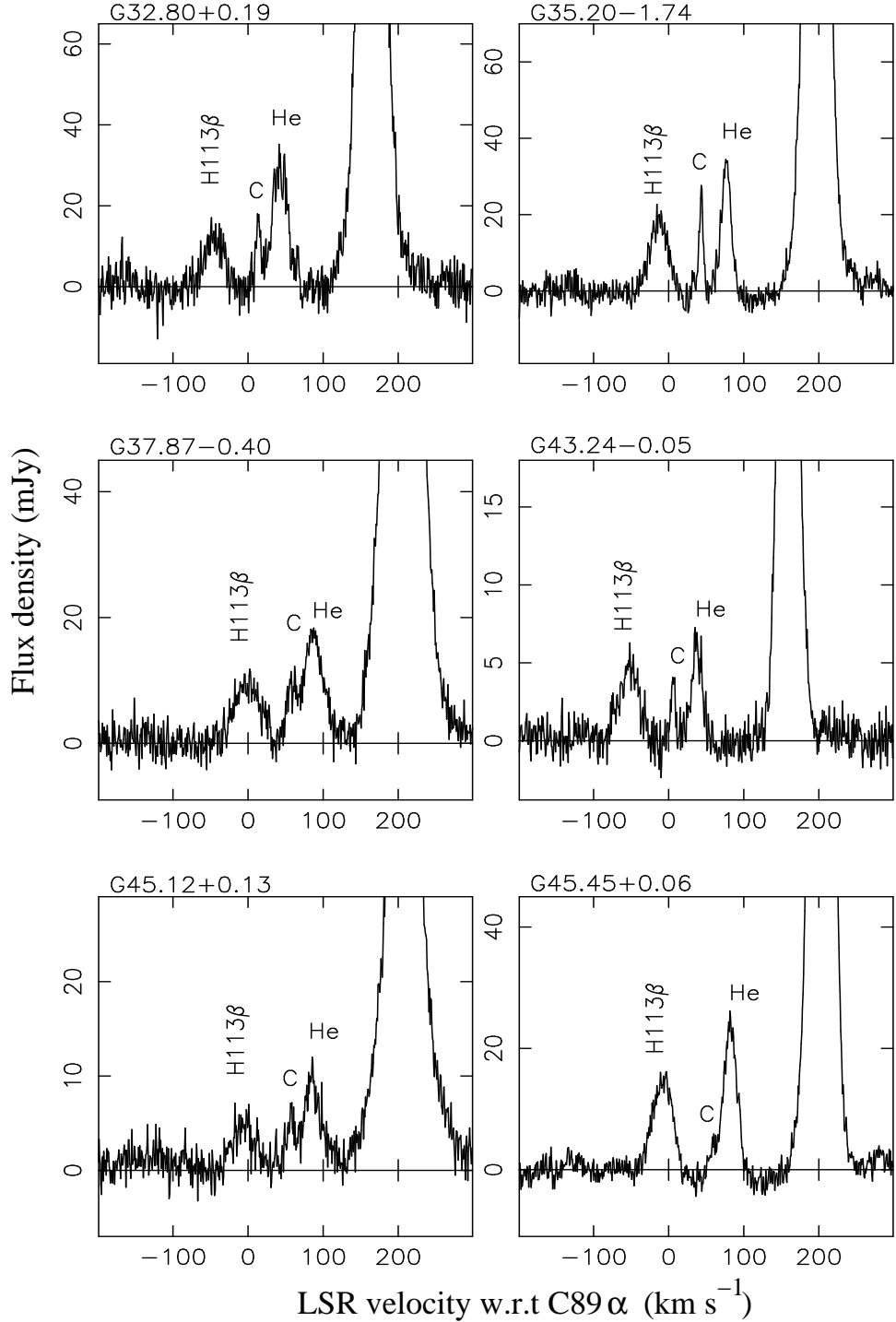}
\caption{Recombination line spectra toward 11 \UCHII s for which
carbon RLs have been detected in
the Arecibo survey. The spectra are obtained by averaging four 
RL transitions (89$\alpha$, 90$\alpha$,
91$\alpha$ and 92$\alpha$) near 8.5 GHz. The strongest line in the figure is the 
hydrogen RL. The carbon, helium, H113$\beta$ and heavy ion RLs are marked
in the spectra where these lines are detected. The spectrum obtained 
toward the `extended' \HII\ region G43.17+0.00 is also shown in the figure.
\label{fig1} }
\end{figure}

Table~\ref{tab1} gives the source names, morphological type and J2000 coordinates  
of the 17 \UCHII s observed in the survey. 
The effective integration time given in Table~\ref{tab1} is eight times the
actual on-source observing time, since the final spectra are obtained by averaging 
all the four RL transitions from two polarizations. The calibrated spectra 
toward the observed sources are shown in Fig.~\ref{fig1}a and \ref{fig1}b.  
Table~\ref{tab2} gives the line parameters obtained from Gaussian modeling 
of the final spectra together with the $1\sigma$ error on these parameters 
provided by the non-linear least square fitting routine. The RMS is obtained
from the final spectrum using values in the velocity range free of any RL 
emission.  The continuum flux densities of the observed sources, which are 
measured with respect to the off-source position used for band-shape measurement, 
are also listed.  This flux density is the mean value of the polynomial baseline 
model removed from the final calibrated spectra (see \S\ref{sec:obs}). Note that 
these values, in all cases, are not the flux densities of the observed sources 
since continuum emission may be present in the off source position. 

As mentioned above, we observed the \HII\ region G43.17+0.0 to test the 
observing setup.  We have detected hydrogen, helium, carbon and H113$\beta$ 
lines toward G43.17+0.0 indicating that our observing procedures were correct.
The coordinates of the \HII\ region and line parameters are given in 
Table~\ref{tab1}.  The final spectrum obtained toward G43.17+0.0 is 
shown in Fig.~\ref{fig1}b. Since G43.17+0.0 is an `extended' \HII\ 
region we will not discuss this source further.   

\subsection{Carbon recombination lines}
\label{sec:crl}

Carbon RLs were detected toward 11 \UCHII s (65\% detection rate) out of the
17 observed sources. The detection of carbon RLs toward a large number of 
\UCHII s indicates that a majority of these \HII\ regions have associated 
dense PDRs.  The sources where carbon RLs were not detected probably also 
have associated dense PDRs;  we believe that the non-detections are due to 
sensitivity limitations.  The ratio of carbon to hydrogen RLs obtained
from the 11 detections has a mean value of 0.11 (1$\sigma$ error 0.09).
The expected carbon line flux density estimated using this ratio and 
the detected hydrogen line flux density is less than the RMS noise in
the spectrum of the sources where carbon RLs were not detected. 
Thus we conclude that our data are consistent with the hypothesis that all
\UCHII s in our survey have associated dense PDR material. 

Based on the continuum morphology of \UCHII s,
\nocite{wc89b}Wood \& Churchwell (1989b) and \nocite{kcw94}Kurtz \etal\ (1994) 
have classified these regions into 7 types: 
unresolved/spherical, cometary, core-halo, Gaussian, 
shell, multiple peak and
irregular. \UCHII s observed in this survey are selected from
these different morphological types (see Table~\ref{tab1}). Carbon
lines have been detected toward \UCHII s belonging to all 
types except for the irregular morphologies.  Observations toward a
larger sample of \UCHII s are, however, needed to establish a
definite relationship between the formation of PDRs and morphology of
\UCHII s.

The observed width of the carbon RLs ranges between 3 and 14 \kms, 
with a median value of 6.3 \kms.  The expected thermal line width 
from these regions is $<$ 1.9 \kms ~(corresponding to a gas temperature 
of 1000 K; \nocite{nwt94}Natta \etal\ 1994), indicating that 
non-thermal motions in the PDR dominate.  
Earlier observations of the NH$_3(1,1)$ transition toward \UCHII s showed 
that their line widths are typically between 1.5 -- 4 \kms, with a median 
value of $\sim$ 3 \kms\ (\nocite{cwc90}Churchwell,
Walmsley \& Cesaroni 1990). Thus the 
median width of carbon RLs is a factor of two larger than that of 
NH$_3(1,1)$.  The line widths in both cases are dominated by non-thermal 
motion. Higher angular resolution observations are needed to understand the 
line width difference and the relationship between the molecular and carbon 
line forming regions. The different line widths may indicate that the two 
lines originate from different regions. Carbon lines may  
originate from regions closer to the boundary of the \UCHII , as expected
in a molecular cloud--\HII\ region interface.

Fig.~\ref{fig2} shows the relationship between the continuum  
and carbon RL flux densities observed toward \UCHII s. Carbon RL intensity
clearly correlates well with the continuum emission. At 8.5 GHz,  
continuum emission originates from the ionized gas associated with the \UCHII\
while carbon RLs originate from PDR gas outside the \HII\ region.   
Thus the correlation between the two observed quantities indicates stimulated
emission of carbon lines by the background continuum radiation. This correlation
confirms the result by \nocite{nwt94}Natta \etal\ (1994) that, 
near 8.5 GHz, carbon RL emission is dominated by stimulated
emission (see also \S\ref{sec:model}). 

The dominance of stimulated emission for carbon lines implies that the
RLs near 8.5 GHz are detected from the PDR on the near side 
of the \UCHII . Further, modeling of RL emission shows that
the line intensity from the PDR at the
far side would typically be a factor of five weaker than that
originating from the front and hence is not detectable in the present
observations (see \S\ref{sec:model}). We used this information 
to study the relative motion between the PDRs and \UCHII s
(\nocite{rgaj05}Roshi \etal\ 2005).  
The observed central velocity
of the hydrogen line represents the mean velocity of the ionized gas
with respect to the observer. Therefore the difference between the
central velocities of the carbon and hydrogen lines provides a measure
of the relative motion of the PDR with respect to the ionized
region. The velocity difference ranges between $-$5.2 and +4.3 \kms,
with an RMS value of 3.3 \kms. Note that the 
mean error of the relative velocity is only 0.4 \kms. Also, the RMS
velocity difference between the hydrogen and carbon RLs 
is 3 times larger than that of hydrogen and helium RLs
(see \S\ref{sec:hy}). Earlier
studies of the relative motion between the ionized gas in compact
\HII\ regions and the molecular gas associated with them have shown
velocity differences ranging between $-$12 and +8 \kms\
(\nocite{fetal90}Forster \etal\ 1990). Our observations show a 
velocity difference that is about half of that inferred from
molecular line data.

\subsection{Hydrogen, Helium, H113\bbeta\ and Heavy ion recombination lines}
\label{sec:hy}

Hydrogen RLs were detected toward all the observed sources. The line profiles
can in most cases be modeled by a single component Gaussian. For 4 sources 
(G34.26+0.15, G37.87$-$0.40, G45.12+0.13 and G70.29+1.60), the observed 
line profile is better approximated by a two component Gaussian (see Table~\ref{tab2}). 
Line width ranges between 23.5 and 69.7 \kms. The broad ($>$ 50 \kms) 
hydrogen lines show non-Gaussian profiles. Such broad lines have been observed earlier 
toward \UCHII s (\nocite{jm99}Jaffe \& Martin-Pintado 1999, 
\nocite{detal04}De Pree \etal\ 2004). Helium and H113$\beta$
RLs were detected toward sources where the RMS noise in the spectrum
is less than about 10\% of the peak hydrogen line flux density. 
G60.88$-$0.13 is an exception -- the H113$\beta$ and carbon line were
detected toward this source but no helium line was detected. A heavy ion
RL was detected toward the source G61.48+0.09.

Since helium and hydrogen RLs originate from the same ionized gas, 
no difference in their central velocities is expected. 
The relative velocity ranges between $-$1.8 and +2.3 \kms\ 
with a RMS velocity difference of 1.1 \kms. Since the mean error of the
relative velocities is 0.6 \kms\ we conclude that this RMS value 
is not significant. (Note that when multiple hydrogen RL components are present
we used the mean velocity of hydrogen RL components 
to compute the relative velocity.) The velocity differences
between the hydrogen and H113$\beta$ RLs show a mean value of
3.6 \kms. The origin of this velocity difference is probably due to the fact
that the H113$\beta$ line is blended with the H129$\gamma$ line.  We have
modeled noiseless Gaussian profiles assuming that the H113$\beta$
transition is 2.5 times more intense than the H129$\gamma$ line, a shift
in LSR velocity of 15 \kms\ between the two line centers (based on the
known frequencies), and that both lines have a FWHM of 25 \kms.  A
single Gaussian component model fit to the blended profile produces a
shift in velocity of 3.67 \kms, consistent with our results.

%The origin of this velocity difference
%could not be definitively traced, but could result from a combination
%of residual baseline structure and blending of the H113$\beta$ line 
%with that of the H129$\gamma$ transition.  

\section{Physical properties of the PDRs}
\label{sec:model}

The physical properties of the PDR can be determined by modeling 
multi-frequency carbon line data (\nocite{nwt94}Natta \etal\ 1994;
\nocite{jetal05}Jeyakumar \etal\ 2005). The observed carbon line intensity 
is a function of the background radiation intensity since the carbon
line emission is dominated by stimulated emission (\nocite{nwt94}Natta \etal\ 1994).
For the present case, the background radiation is due to the thermal
continuum emission from the \UCHII s. Therefore, for modeling, 
we carefully selected a subset of \UCHII s from our sample
whose properties could be estimated with data available in the
literature. Table~\ref{tab3} summarizes the parameters of the 
\UCHII s selected for modeling. 
Interferometric images of the selected sources near 3.6 cm wavelength
with visibilities measured at angular scale comparable to the
Arecibo beam are available (\nocite{aetal96}Afflerbach \etal\ 1996; 
\nocite{kcw94}Kurtz \etal\ 1994). Continuum flux 
density and angular size of the \UCHII s are obtained from these images.
Note that for sources where both compact and `extended' emission are present
the angular size is roughly obtained from the half-power contour 
of the extended emission. Thus these source sizes are different from the
values given by \nocite{aetal96}Afflerbach \etal\ (1996) in their paper.
\nocite{aetal96}Afflerbach \etal\ (1996) have estimated
the electron temperature, $T_e$, of the ionized gas in most of the selected \UCHII s.
We assumed a temperature of 10$^4$ K for the source G43.24$-$0.05 since no
temperature estimate is available. Using these temperatures and angular sizes, 
we derive the emission measures, $EM$s, of the \UCHII s that are consistent 
with the observed continuum flux density at 3.6 cm. We assume that the ionized 
region has cylindrical geometry with length equal to the diameter
to derive the $EM$. (The correction factor
given by \nocite{mh67}Mezger \& Henderson (1967) that needs to 
be used while obtaining the $EM$ for such geometry
is not applied since for most sources the flux density distribution cannot be
approximated as a simple Gaussian.) 
These $EM$s along with angular sizes and distances to the sources
are used to estimate the electron density, $n_e$.

The PDR parameters that determine the carbon line intensity are
gas temperature, $T_{PDR}$, electron density, $n_e^{PDR}$, and PDR thickness along
the line of sight, $l$. We considered homogeneous slabs of carbon line forming region 
in front and back of the \UCHII s to estimate the RL emission. The line
brightness temperature, $T_{LB}$, due to the slab in the front side is given by
(\nocite{s75}Shaver 1975)
\begin{eqnarray}
T_{LB} & = & T_{0bg,\nu} e^{-\tau_{C\nu}}(e^{-\tau_{L\nu}} - 1)\nonumber \\
    &   & \mbox{} +  T_{PDR}\left( \frac{b_m \tau_{L\nu}^* + \tau_{C\nu}}{\tau_{L\nu} + \tau_{C\nu}}
          (1 - e^{-(\tau_{L\nu} + \tau_{C\nu})}) - (1 - e^{-\tau_{C\nu}})\right) 
\label{eq:line}
\end{eqnarray}
where $T_{0bg,\nu}$ is the background radiation temperature and  
$\tau_{C\nu}$ is the continuum optical depth of the PDR. The non-LTE line 
optical depth of the spectral transition from energy state $m$ to $n$, 
$\tau_{L\nu}$, is $\tau_{L\nu}  =  b_n \beta_n \tau_{L\nu}^*$
where $\tau_{L\nu}^*$ is the LTE line optical depth,  
$b_n$ and $\beta_n$ are the departure coefficients of state $n$. For the
PDR, $\tau_{L\nu}^*\; \propto\;  n_e^{PDR} n_{C^+} l$, where $n_{C^+}$ is the carbon
ion volume number density in the PDR. For the present calculations we assumed 
that $n_e^{PDR} = n_{C^+} = n_e$ so $\tau_{L\nu}^*\; \propto\;  n_e^2 l$.
Thus, for a homogeneous PDR, the electron density provided by the modeling will
be the true electron density. However, for an inhomogeneous PDR,
the actual electron density depends on just how clumpy the PDR is and
the model value will be the square root of the second moment of the 
spatial distribution of electron density. 
The background temperature is given by
\begin{equation}
T_{0bg,\nu} = T_e (1 - e^{-\tau_{C\nu}^{UCH}}) \; ,
\end{equation}
where $\tau_{C\nu}^{UCH}$ is the continuum optical depth of the \UCHII . The line
temperature from the PDR on the far (back) side is obtained from Eq.~\ref{eq:line}
by setting $T_{0bg,\nu} = 0$. The line brightness temperature is
finally converted to flux density using the source angular size given
in Table~\ref{tab3}. 

Modeling starts with the computation of 
the non-LTE departure coefficients for a given 
electron density, gas temperature and background radiation 
field. The coefficients are obtained using the program developed originally by 
\nocite{bs77} Brocklehurst \& Salem (1977) and later modified by
\nocite{pae94} Payne, Anantharamaiah \& Erickson (1994). 
These coefficients are then used to estimate the carbon line flux density
as a function of frequency. We combined our data at 8.5 GHz with the 
existing RL data at 4.8 GHz to constrain the electron density and 
PDR thickness for a given gas temperature. The 4.8 GHz data 
(\nocite{aetal02}Araya \etal\ 2002; \nocite{wetal03}Watson \etal\ 2003)
are listed in Table~\ref{tab3}. We found that, for a given temperature, 
the electron density and PDR thickness can be well constrained using the
data at these two frequencies.  Particularly, the upper limit near 4.8 GHz provides 
a stringent lower limit on the electron density in the PDR (see also \S\ref{sec:rob}). 

Typically the PDR gas temperature is in the range 300 -- 1000 K 
(\nocite{nwt94}Natta \etal\ 1994). Table~\ref{tab4} lists the PDR electron 
density and thickness for temperatures 300 K and 500 K. Note that for higher 
gas temperature the estimated electron density will be larger. The carbon 
RL flux densities predicted by the Table~\ref{tab4} models are shown in 
Fig.~\ref{fig3} which plots for each source the model carbon flux density as a 
function of frequency. 
The model electron densities are converted to neutral densities using the cosmic 
abundance of carbon (3.9 $\times$ 10$^{-4}$; \nocite{m74}Morton 1974) and a 
depletion factor of 25\% (\nocite{nwt94}Natta \etal\ 1994). The neutral density, 
$n_H$ (number density 
of hydrogen atom), is in most cases greater than 5 $\times$ 10$^5$ \cmthree 
(see Table~\ref{tab4}). 

An important inference from modeling is the dominance of stimulated emission
for carbon lines at frequencies \lsim\ 15 GHz. Line emission
from the PDR slab in front of the \UCHII\  is amplified
by the background continuum radiation from the ionized gas. 
The line flux density is typically 5 times larger than that 
from the PDR slab behind the \UCHII . Thus
the sensitivity of our observations
is sufficient to detect only carbon line emission from
the front of the \UCHII . As discussed in \S\ref{sec:crl}, 
this fact can be used to study the relative motion of 
the PDR  with respect to the \UCHII .

\subsection{Pressure}

We use the derived physical properties of the PDRs and the \UCHII s
to estimate the pressure in these regions. 
The total gas pressure inside the \UCHII\  is the sum of thermal 
pressure and turbulent pressure. Because there are no direct 
measurements of the magnetic field strength in \UCHII s or PDRs 
we do not include any contribution to the pressure from magnetic fields.
%Magnetic fields inside \HII\ regions are typically $\sim$ 10 -- 20 $\mu$G 
%(e.g. \nocite{hc80}Heiles \& Chu 1980) and hence do not contribute 
%significantly to the total pressure. 
The total gas pressure is given by
\begin{equation}
P_{UCH} = 2 k n_e T_e + n_e \mu m_H v_{turb}^2 \mbox{~~dyne cm$^{-2}$},
\label{eq:pres}
\end{equation}
where $k$ is the Boltzmann constant, $\mu = 1.4$ is the effective mass
in amu of a pure H + He gas with He fraction taken as 10\% by number,
$m_H$ is the hydrogen mass in gm and $v_{turb}$ is 
the turbulent velocity inside the \UCHII\  in units of cm s$^{-1}$. 
The line profile due to turbulence is considered to be
Gaussian with an FWHM of $v_{turb}$. We estimated $v_{turb}$ from the 
observed width of helium lines after removing the contribution due to
thermal motion. (Hydrogen lines were not used since
some of the \UCHII s have multiple line components.) The values for
electron density and temperature are taken from Table~\ref{tab3}. The
estimated pressures of the ionized gas in \UCHII s are typically $<$
2 $\times$ 10$^{-7}$ \dycm (see Table~\ref{tab3}). Our calculations
show that the turbulent pressure in \UCHII s is about an order of magnitude
larger than the thermal pressure. 

The total gas pressure in the PDR is given by
\be
P_{PDR} = k n_H T_{PDR} + n_H \mu m_H v_{C-turb}^2 \mbox{~~\dycm},
\ee
where $v_{C-turb}$ is the turbulent velocity 
in cm s$^{-1}$ estimated from the FWHM of the observed carbon RLs after removing
the thermal line width. To estimate 
PDR pressure we consider that in the region where carbon is 
ionized, the neutral material is predominantly in atomic hydrogen form
(see \nocite{ht97}Hollenbach \& Teilens 1997). 
%The value of magnetic field strength in the PDR is not known and hence
%we do not include its contribution in the calculation of PDR pressure. 
The estimated total gas pressure in the PDR is typically  10$^{-6}$ \dycm
(see Table~\ref{tab4}). As in the case of the \UCHII , the turbulent 
pressure in the PDR is about one order of magnitude larger than the thermal pressure.
This importance of turbulent pressure was noted earlier by 
\nocite{xetal96}Xie \etal\ (1996).

\subsection{Robustness of the PDR parameters and Pressure estimates}
\label{sec:rob}

For modeling the PDR properties we have used the carbon line parameters
obtained from our survey. However, as discussed in \S\ref{sec:obs},
the line flux densities are uncertain by up to 20\%. To study
the effect of this uncertainty, we
estimated the physical properties of the PDR by changing the 
line flux density near 8.5 GHz by $\pm$ 20\% and 
compared them with the values given in 
Table~\ref{tab4}. We found that the estimated neutral densities
and PDR thickness change by about a factor of two and three
respectively. Fig.~\ref{fig4} shows the result of such a study
toward G32.80+0.19. The PDR parameters and pressures obtained
in this study are given in Table~\ref{tab5}.
%line flux density was increased by 20\% for $T_{PDR} = 300$ K
%are $n_e^{PDR} = 1000$ \cmthree, $l = $ 6.2 $\times$ 10$^{-4}$ pc
%and those for $T_{PDR} = 500$ K are $n_e^{PDR} = 2000$ \cmthree, 
%$l = $ 4.5 $\times$ 10$^{-4}$ pc. When the line flux density is
%reduced by 20\% the parameters obtained are $n_e^{PDR} = 300$ \cmthree, 
%$l = $ 2.6 $\times$ 10$^{-3}$ pc and $n_e^{PDR} = 400$ \cmthree,
%$l = $ 3.6 $\times$ 10$^{-3}$ pc for gas temperatures of 300 K
%and 500 K respectively. 
We conclude that the
lower limits on neutral densities and pressure of the PDR given
in Table~\ref{tab4} will reduce only by about a factor of two 
due to the measurement uncertainty of the carbon RL flux 
density at 8.5 GHz.   

\section{Discussion}
\label{sec:con}

A survey of recombination lines near 8.5 GHz toward 17 \UCHII s
using the Arecibo telescope detected carbon lines from 11 sources (65\% 
detection rate). In this paper, we have shown that the carbon RLs provide
important, complementary information on the kinematics and properties
of the ambient medium near \UCHII s. The RLs were used 
to study the relative motion between the carbon line forming region 
and the \UCHII s. We obtained a RMS velocity difference of 3.3 \kms. 
Earlier, \nocite{fetal90}Forster \etal\ (1990) compared hydrogen RL and 
molecular line velocities toward nine compact \HII\ regions to 
study their relative motion. They concluded that the RL velocities
are offset with respect to molecular lines by typically $\pm$ 6 \kms. 
The relative velocity difference obtained from our observations is about
a factor of two smaller than that measured by \nocite{fetal90}Forster \etal\ (1990).
These authors argue that the larger velocity difference they have measured is 
consistent with the expansion of a 10$^4$ K \HII\ region at 
the sound speed of $\sim$ 15 \kms. The lower velocity difference 
seen in our observations may
indicate that \UCHII s are not expanding at sound speeds in
ionized gas.  

The dynamical lifetime for \UCHII s of a few times  $10^3$ years 
was estimated assuming that the \HII\ region is
expanding at the speed of sound in the ionized gas. 
However, the lifetime deduced from the observed number of \UCHII s
is a few times  $10^5$ years.
This discrepancy between the two lifetimes is referred to
as the ``lifetime problem'' of \UCHII s
(\nocite{wc89a}Wood \& Churchwell 1989a). Several hypothesis
have been proposed to explain the long lifetime of \UCHII s
(see \nocite{c99}Churchwell 1999).  For example,
\nocite{drg95}De Pree, Rodr\'iguez \& Goss (1995)
suggested that if high density ($\sim$ 10$^7$ \cmthree), 
warm ($\sim$ 100 K) molecular material is present 
in the vicinity of \UCHII s, it may be able to pressure confine \UCHII s 
that form there and thus extend their lifetime.
This hypothesis is supported by
recent molecular line observations (eg. \nocite{ketal00}Kurtz \etal\ 2000),
which show that the temperature  and density of the
molecular cloud harboring \UCHII s are $>$ 50 K and $\sim$ 10$^7$ \cmthree
respectively. Modeling of carbon RL data in our observations also indicates that 
\UCHII s are embedded in neutral regions with high ambient pressure.

In the pressure-confined nebula scenario, if the massive
star is stationary with respect to the molecular core, then
no relative motion between the ionized gas and the PDR is expected. 
However, our carbon line observations indicate a non-zero relative motion
between the PDR and the ionized gas (see \S\ref{sec:crl}). 
If the star is moving with respect to the cloud core, such relative
motion is expected even though the ionized gas is pressure confined. 
The star motion can also produce the different observed
morphologies of \UCHII s (\nocite{gf04}Garcia-Segura \& Franco 2004). 
However, relative motion between the PDR and ionized gas is expected 
in other models proposed to solve 
the long lifetime of \UCHII s and their morphology. For example, in 
the model proposed by \nocite{kk01}Kim \& Koo (2001), the ionizing
star is at the edge of a dense, hot core and the `champagne flow' through the
hierarchical structure of the molecular cloud produces the extended
emission that has been observed around \UCHII s. The PDR, in this model,
resides in the dense, hot core. The presence of the `champagne flow' can shift 
the centroid of the RL emission from the ionized gas relative to the 
line emission from the PDR. This shift results in a velocity difference 
between the hydrogen and carbon RLs from the \UCHII s in our observations. 
Higher angular resolution observations of carbon RLs may provide
enough information to distinguish between these models.

Interestingly, the PDR pressure we derive, in most cases, 
is an order of magnitude larger than the total pressure in \UCHII s. 
In the classical treatment of the pressure driven evolution
of \UCHII s, a dense shell of neutral material is formed 
surrounding the \HII\ region 
(\nocite{gf96}Garcia-Segura \& Franco 1996). If 
the PDR responsible for the carbon line emission 
resides in this neutral shell then its pressure
can be high. However, if ionization front is trapped within the neutral
material for $\sim$ 10$^5$ years, then the sound crossing 
length scale ($\sim$ 0.2 pc; sound speed $\sim$ 2 \kms)
is larger than the typical molecular cloud core sizes ($\le$ 0.1 pc;
Churchwell 2002). Thus the neutral shell would have diffused out
in a fraction of the expected lifetime of the \UCHII s and hence
the ambient pressure outside the \UCHII\ is most likely that of the natal
molecular cloud. The large pressure difference, therefore, may 
indicate that the dominant pressure inside the \UCHII\  
is due to other physical processes such as stellar wind. Further investigation 
with high angular resolution, multi-frequency RL and continuum observations 
together with modeling of the evolution
of \HII\ regions surrounded by a dense PDR is needed to understand this 
large pressure difference. 

\acknowledgments

We are grateful to Phil Perillat (NAIC) for his valuable assistance 
during observations and for his analysis software.
We thank Chris Salter (NAIC) for 
his generous help in processing the calibration data 
and the telescope operators for their support during the observations.

%Observed sources

\begin{deluxetable}{lllrr}
%\tabletypesize{\small}
\tablecolumns{5}
\tablewidth{0pc}
\tablecaption{Summary of Arecibo survey
\label{tab1}}
\tablehead{
\colhead{Source} & \colhead{Type\tablenotemark{a}} & \colhead{RA(2000)} & \colhead{DEC(2000)} & \colhead{Integration time}  \\
\colhead{Name} &  & \colhead{(hh:mm:ss.s)} & \colhead{(dd:mm:ss)} & \colhead{(minutes)} }
\startdata
G32.80+0.19 & \nodata\tablenotemark{b} &   18:50:30.9 &  $-$00:02:00 & 40   \\
G34.26+0.15 & \nodata\tablenotemark{b} &   18:53:18.5 &  +01:14:59   & 40   \\
G35.20$-$1.74 & C\tablenotemark{c}     &   19:01:46.5 &  +01:13:24   & 80  \\
G35.57$-$0.03 & \nodata\tablenotemark{b}&   18:56:22.6&  +02:20:27   & 40   \\
G37.87$-$0.40 & C                      &   19:01:53.6 &  +04:12:49   & 80  \\
G41.74+0.10 & CH                       &   19:07:15.5 &  +07:52:44   & 40   \\
G43.24$-$0.05& \nodata\tablenotemark{b}&   19:10:33.5 &  +09:08:25   & 320  \\
G43.89$-$0.78 & C                      &   19:14:26.2 &  +09:22:34   & 40   \\
G45.12+0.13 & C                        &   19:13:27.8 &  +10:53:37   & 120  \\
G45.45+0.06 & C                        &   19:14:21.4 &  +11:09:15   & 160  \\
G45.47+0.05 & I                        &   19:14:25.6 &  +11:09:26   & 40   \\
G48.61+0.02 & \nodata\tablenotemark{b} &   19:20:31.2 &  +13:55:23   & 120   \\
G50.32+0.68 & S                        &   19:21:27.6 &  +15:44:21   & 40    \\
G60.88$-$0.13 & G                      &   19:46:19.9 &  +24:35:24   & 240   \\
G61.48+0.09& S                        &   19:46:48.3 &  +25:12:48   & 160   \\
G70.29+1.60 & CH                       &   20:01:45.7 &  +33:32:43   & 120   \\
G70.33+1.59 & U                        &   20:01:54.1 &  +33:34:15   & 120   \\ \hline
G43.17+0.00\tablenotemark{d} & \nodata  &   19:10:15.7 &  +09:06:06   & 40    \\
\enddata
\tablenotetext{a}{Morphological type of \UCHII s: C -- cometary, CH -- core-halo,
                  I -- irregular, S -- spherical, G -- Gaussian, U -- unresolved. The
                  type of the observed \UCHII s is taken from Wood \& Churchwell
                  (1989) and Kurtz \etal\ (1994)}    
\tablenotetext{b}{Multiple sources with different morphological types are present within the
                  observed beam.}
\tablenotetext{c}{An evolved \HII\ region is present within the observed beam.}
\tablenotetext{d}{`Extended' \HII\ region (Lockman 1989)}
\end{deluxetable}

%Parameters of observed lines 

\begin{deluxetable}{lrrlrrr}
\tabletypesize{\small}
\tablecolumns{7}
\tablewidth{0pc}
\tablecaption{Parameters of the observed recombination lines
\label{tab2}}
\tablehead{
\colhead{Source} & \colhead{$\Delta S\tablenotemark{a}$} & \colhead{RMS} & \colhead{Transition} & \colhead{$S_L$} & \colhead{$V_{LSR}$} & \colhead{$\Delta V$}   \\
\colhead{Name}  & (Jy)    & (mJy) &                 & (mJy) & (\kms) & (\kms) } 
\startdata
G32.80+0.19 & 3.8 & 3.5 & C          &  16.2(2.4) & 13.5(0.6)  & 7.5(1.3)   \\ 
            &     &     & He         &  29.6(1.3) & 15.1(0.6)  & 24.3(1.3)  \\
            &     &     &  H          & 326.5(1.2) & 15.3(0.1) & 30.1(0.1)  \\ 
            &     &     &  H113$\beta$& 57.4(1.8) & 12.2(0.5) & 35.4(1.3)  \\
G34.26+0.15 & 3.5 & 2.9 & C          &\nodata     & \nodata    & \nodata    \\  
            &     &     & He         &   7.9(0.6) & 41.7(1.9)  & 53.2(4.4)  \\
            &     &     & H          &  47.2(1.9) & 51.0(0.3) & 26.5(1.0)  \\ 
            &     &     & H          &  85.4(1.7) & 36.0(0.4) & 64.9(0.6)  \\ 
            &     &     & H113$\beta$&\nodata     & \nodata    & \nodata    \\
G35.20$-$1.74 & 4.1 & 2.6 & C          &  25.2(2.0) & 43.3(0.3)  & 6.3(0.6)  \\
            &     &       &He         &  33.4(1.3) & 49.2(0.3)  & 15.1(0.7)  \\
            &     &       &H          &  372.0(1.0)& 47.8(0.04)& 27.5(0.1) \\
            &     &       &H113$\beta$& 81.9(1.6) & 45.3(0.3) & 30.1(0.7)  \\
G35.57$-$0.03 & 0.4 & 2.3 & C          &  \nodata   & \nodata    & \nodata   \\
            &     &       & He         &  \nodata   & \nodata    & \nodata    \\
            &     &       & H          &  22.0(0.6) & 52.3(0.4) & 30.9(0.9) \\ 
            &     &       & H113$\beta$&  \nodata   & \nodata    & \nodata    \\ 
G37.87$-$0.40 & 3.0 & 2.1 & C          &  8.0(0.7)  & 56.8(0.6)  & 14.0(1.5) \\
            &     &       & He         &  16.8(0.4) & 59.7(0.5)  & 31.7(1.3)  \\
            &     &       & H          &  93.0(2.7) & 60.3(0.1) & 26.1(0.4)  \\
            &     &       & H          &  104.5(2.7)& 56.6(0.1)& 55.1(0.5)  \\
            &     &       & H113$\beta$& 42.7(1.3) & 56.4(0.6) & 38.7(1.4)  \\
G41.74+0.10 & 0.1 & 1.8   & C          &  \nodata   & \nodata    & \nodata   \\
            &     &       & He         &  \nodata   & \nodata    & \nodata    \\
            &     &       & H          &  7.7(0.5)  & 12.9(1.0) & 34.0(2.3) \\
            &     &       & H113$\beta$&  \nodata   & \nodata    & \nodata    \\
G43.24$-$0.05& 0.7 & 0.8  & C          &3.9(0.5)    & 5.9(0.4)   & 6.2(0.9)    \\
            &     &       & He         &6.0(0.3)    & 10.0(0.4)  & 15.8(0.9)   \\
            &     &       & H          & 73.3(0.2)  & 9.6(0.04)& 24.1(0.1) \\
            &     &       & H113$\beta$& 18.2(0.6) &  5.7(0.5) & 29.9(1.2) \\
G43.89$-$0.78 & 0.2 & 1.9 & C          & \nodata   & \nodata    & \nodata   \\
            &     &       & He         & \nodata   & \nodata    & \nodata    \\
            &     &       & H          & 18.4(0.5) & 54.4(0.4) & 25.9(0.8) \\
            &     &       & H113$\beta$& \nodata   & \nodata    & \nodata    \\ 
G45.12+0.13 & 1.9 & 1.3   & C          & 4.9(0.6)  & 55.7(0.8)  & 12.9(1.9) \\
            &     &       & He         & 8.9(0.4)  & 58.5(0.7)  & 30.1(1.8)  \\
            &     &       & H          & 72.5(1.2) & 57.1(0.1) & 27.4(0.4) \\
            &     &       & H          & 32.7(1.3) & 59.7(0.3) & 69.5(1.2) \\
            &     &       & H113$\beta$& 21.9(2.2) & 51.8(1.4) & 27.6(3.2) \\
G45.45+0.06 & 2.6 & 1.6   &  C         &  3.5(0.8) & 59.0(1.2)  & 9.6(2.9)  \\
            &     &       & He        &  23.9(0.6) & 54.6(0.3)  & 20.3(0.7)  \\
            &     &       & H         & 248.8(0.5)& 54.7(0.03)& 27.6(0.06) \\
            &     &       & H113$\beta$& 58.4(2.5) & 51.3(0.7) & 33.0(1.6)  \\
G45.47+0.05 & 0.2 & 2.2   & C          & \nodata   & \nodata    & \nodata   \\
            &     &       & He         & \nodata   & \nodata    & \nodata    \\
            &     &       & H          & 12.8(0.5) & 62.9(0.6) & 31.6(1.5) \\
            &     &       & H113$\beta$& \nodata   & \nodata    & \nodata    \\  
G48.61+0.02 & 0.5 & 1.2   & C          &  5.2(1.0) & 18.4(0.3)  & 3.3(0.7)  \\
            &     &       & He         &  4.5(0.5) & 15.8(0.7)  & 11.7(1.6)  \\
            &     &       & H          &  50.5(0.4)& 16.7(0.1) & 23.5(0.2) \\
            &     &       & H113$\beta$& 18.1(1.1) & 12.4(0.9) & 30.5(2.2) \\
G50.32+0.68 & 0.1 & 2.0   & C          & \nodata   & \nodata    & \nodata   \\
            &     &       & He         & \nodata   & \nodata    & \nodata    \\
            &     &       & H          & 12.0(0.5) & 26.9(0.5) & 25.7(1.2) \\
            &     &       & H113$\beta$& \nodata   & \nodata    & \nodata   \\
G60.88$-$0.13 & 0.2 & 1.0 & C          &  11.5(0.7)& 22.0(0.1)  & 4.5(0.3)  \\
            &     &       & He         & \nodata   & \nodata    & \nodata    \\
            &     &       & H          & 35.2(0.3) & 18.3(0.1) &  20.7(0.2) \\
            &     &       & H113$\beta$&  8.9(0.8) & 12.7(1.4) & 30.6(3.3) \\
G61.48+0.09& 4.0 & 1.6   & C          &  45.5(1.7)& 21.3(0.1)  &  4.7(0.2) \\
            &     &       & He         &  27.2(0.9)& 28.8(0.3)  & 16.1(0.6) \\ 
            &     &       & H          & 368.5(0.7)& 26.4(0.03)& 26.3(0.1) \\
            &     &       & H113$\beta$& 78.0(3.7) & 23.2(0.8) & 33.4(1.8)  \\
            &     &       & S(?)\tablenotemark{b}     &  7.2(2.4) & 21.3(0.4) & 2.4(0.9)   \\
G70.29+1.60 & 5.3 & 1.6   & C          & 24.5(1.3) & $-$25.6(0.2) & 8.3(0.5)  \\
            &     &       & He         & 17.6(0.5) & $-$27.4 (1.0) & 62.1(2.2)  \\
            &     &       & H          & 132.6(2.1)& $-$27.1(0.1) & 26.2(0.3) \\
            &     &       & H          & 130.1(2.1)& $-$28.8(0.1) &  69.7(0.6) \\
            &     &       & H113$\beta$& 41.3(1.7) & $-$27.8(0.7) & 34.9(1.6) \\
G70.33+1.59& 2.9 & 1.8   & C          & 19.9(1.6) & $-$21.6(0.2) & 4.2(0.4)  \\
            &     &       & He         & 18.5(0.7) & $-$18.8(0.4)  & 22.6(1.0)  \\
            &     &       & H          & 222.5(0.6)& $-$19.6(0.04)& 29.6(0.1) \\
            &     &       & H113$\beta$& 59.4(1.2) & $-$23.5(0.3) & 33.3(0.7) \\ 
\cutinhead{`Extended' \HII\ region}
G43.17+0.0  & 9.0 & 3.9   & C          &  13.1(2.1)& $-$28.4(1.3) & 14.1(3.1) \\
            &     &       & He         & 76.9(1.7) & $-$30.2(0.3)  & 23.0(0.8) \\
            &     &       & H          &  774.6(..)& $-$30.0(0.03)& 29.9(..)     \\
            &     &       & H113$\beta$& 206.0(2.6)& $-$33.6(0.2) & 33.1(0.5)  \\
\enddata
\tablenotetext{a}{Continuum flux density measured with respect to 
                  the off-source position used for bandpass calibration.
                  The error in continuum flux density is 20\% (see text).}
\tablenotetext{b}{Presumably a heavy ion (S$^+$ ?) RL. While obtaining the parameters of 
                  this RL using the fitting routine we kept the parameters of 
                  the other four RLs constant.}
\end{deluxetable}

%Parameters used for modeling
%
\begin{deluxetable}{lrllrrrcr}
\tabletypesize{\small}
\tablecolumns{9}
\tablewidth{0pc}
\tablecaption{Input parameters for modeling 
\label{tab3}}
\tablehead{
\colhead{Source} & \colhead{D} & \colhead{S$_{3.6}$\tablenotemark{a}} & \colhead{S$_{C110\alpha}$\tablenotemark{b}} & \colhead{Size\tablenotemark{a}} & \colhead{T$_e$ } & \colhead{$EM$} &  \colhead{P$_{UCH}$\tablenotemark{c}} & \colhead{Ref\tablenotemark{~d}} \\
\colhead{Name}   & \colhead{(kpc)} & \colhead{(Jy)} & \colhead{(mJy)} & \colhead{(\arcsec $\times$ \arcsec)} & \colhead{(K)} & \colhead{(pc cm$^{-6}$)} & \colhead{(10$^{-8}$ \dycm)} & \colhead{d,r,s,t} 
}
\startdata
G32.80+0.19 & 13.0 & 3.6 & $<$ 5.0  &  15 $\times$ 7 & 7600\tablenotemark{e} & 17$\times 10^6$ &  7.1  & 1,4,1,1 \\
G37.87$-$0.40 & 9.2 & 4.2 & $<$ 5.0  & 10 $\times$ 10 & 7800 & 21$\times 10^6$ & 16.1 & 1,4,1,1  \\    
G43.24$-$0.05& 11.6 & 0.4 & $<$ 1.4  & 8 $\times$ 5 & 10000\tablenotemark{f} & 5$\times 10^6$ & 2.2  & 2,5,6,..  \\ 
G45.12+0.13 & 6.0 & 6.1 & $<$ 6.0  & 15 $\times$ 7 & 8300 & 31$\times 10^6$ &  21.6 & 3,4,1,1 \\ 
G45.45+0.06 & 7.4 & 4.6 & $<$ 6.0  & 20 $\times$ 20 & 9800 & 6$\times 10^6$ & 2.8 & 1,4,1,1 \\    
G70.29+1.60 & 8.6 & 5.0 & $<$ 10.0 &  10 $\times$ 5 & 8500 & 56$\times 10^6$ & 123.8 & 1,4,1,1 \\
G70.33+1.59& 8.6 & 1.6 & $<$ 10.0 &  15 $\times$ 7 & 9100 & 8$\times 10^6$ & 5.2 & 1,4,1,1 \\   
\enddata
\tablenotetext{a}{3.6 cm continuum flux density and angular size are obtained from interferometric images}
\tablenotetext{b}{C110$\alpha$ (4876.589 MHz) flux densities are taken from the literature}
\tablenotetext{c}{The total pressure in \UCHII\ (see text)}
\tablenotetext{d}{References from which distance (d), 4.8 GHz line flux density (r),
                 angular size of \UCHII s (s) and electron temperature (t) 
                 were obtained.}
\tablenotetext{e}{An average electron temperature of the two \UCHII s, 
                  G32.80+0.19A \& G32.80+0.19B, present within the observed beam 
                  were obtained from Ref (1)}
\tablenotetext{f}{An electron temperature of 10$^4$ K was assumed for modeling}
\tablerefs{(1)Afflerbach \etal\ (1996); (2)Kurtz \etal\ (1999); (3)Fish \etal\ (2003);
           (4)Araya \etal\ (2002); (5)Watson \etal\ (2003); (6)Kurtz \etal\ (1994). }
\end{deluxetable}

%Output of modeling
\begin{deluxetable}{lrrccrrcc}
\tabletypesize{\small}
\tablecolumns{9}
\tablewidth{0pc}
\tablecaption{Results of modeling 
\label{tab4}}
\tablehead{
\colhead{Source} & \multicolumn{4}{c}{\underline{~~~~~~~~~~~~~PDR\tablenotemark{a} with $T_{PDR} = 300$ K ~~~~~~~~~~~~~}} & \multicolumn{4}{c}{\underline{~~~~~~~~~~~~~PDR\tablenotemark{a} with $T_{PDR} = 500$ K ~~~~~~~~~~~~~}} \\
\colhead{Name} & $n_e^{PDR}$ & $l$ & $n_H$ & \colhead{P$_{PDR}$} & $n_e^{PDR}$ & $l$ & $n_H$ & \colhead{P$_{PDR}$}   \\  
    & \colhead{(cm$^{-3}$)} & \colhead{(pc)} & \colhead{(cm$^{-3}$)} & \colhead{(\dycm)} & \colhead{(cm$^{-3}$)} & \colhead{(pc)} & \colhead{(cm$^{-3}$)} & \colhead{(\dycm)}  \\ 
  &  &\colhead{$\times$ 10$^{-3}$}  & \colhead{$\times$ 10$^6$}  & \colhead{$\times$ 10$^{-6}$} &  &\colhead{$\times$ 10$^{-3}$} &\colhead{$\times$ 10$^{6}$}  & \colhead{$\times$ 10$^{-6}$}} 
\startdata
G32.80+0.19 & 500 & 1.4 &1.7  &2.2 & 1000 & 1.1 &3.4  &4.5         \\
G37.87$-$0.40 & 110 & 15.0 & 0.4  & 1.7  & 120 & 29.0 & 0.4  & 1.8  \\
G43.24$-$0.05& 300 & 4.0 & 1.0  & 0.9  & 300 & 9.0 & 1.0  & 0.9     \\ 
G45.12+0.13 & 40 & 48.0 & 0.1  & 0.5  & 40 & 124.0 & 0.1  & 0.5      \\
G45.45+0.06 & 20 & 187.0 & 0.1  & 0.2  & 30 & 150.0 & 0.1 & 0.2     \\
G70.29+1.60 & 300 & 4.8 & 1.0  & 1.6  & 350 & 8.6 & 1.2  & 1.9      \\
G70.33+1.59& 150 & 13.5 & 0.5 & 0.2  & 170 & 25.0 & 0.6  & 0.3     \\
\enddata
\tablenotetext{a}{The electron density, neutral density and PDR pressure are lower limits}
\end{deluxetable}

%Robustness of modeling
\begin{deluxetable}{ccrccccc}
\tabletypesize{\small}
\tablecolumns{8}
\tablewidth{0pc}
\tablecaption{Robustness of the estimated parameters 
\label{tab5}}
\tablehead{
\multicolumn{8}{c}{Source G32.80+0.19\tablenotemark{a}} \\\hline
\colhead{$T_{PDR}$} & \colhead{S$_L$} & \colhead{$n_e^{PDR}$} & \colhead{ $l$} & \colhead{$n_H$} & \colhead{$n_H$/$n_H^{'}$\tablenotemark{b}}  & \colhead{P$_{PDR}$} & \colhead{P$_{PDR}$/P$_{PDR}^{'}$\tablenotemark{b}} \\
\colhead{(K)} & \colhead{(mJy)} & \colhead{(cm$^{-3}$)} & \colhead{(pc)} & \colhead{(cm$^{-3}$)} &  & \colhead{(\dycm)} &   \\ 
  &  & & \colhead{$\times$ 10$^{-3}$}  & \colhead{$\times$ 10$^6$} & & \colhead{$\times$ 10$^{-6}$}& } 
\startdata
300 & 13.0\tablenotemark{c} & 300  & 2.6  & 1.0  & 0.6 & 1.3  & 0.6 \\
300 & 19.4\tablenotemark{d} & 1000  & 0.6 & 3.4  & 2.0 & 4.4 &  2.0 \\
500 & 13.0\tablenotemark{c} & 400  & 3.6  & 1.3  & 0.4 & 1.8  & 0.4 \\
500 & 19.4\tablenotemark{d} & 2000 & 0.5  & 6.7  & 2.0 & 8.9 &  2.0 \\
\enddata
\tablenotetext{a}{The electron density, neutral density and PDR pressure are lower limits.}
\tablenotetext{b}{Ratio of the neutral density and PDR pressure estimated with $\pm$ 20\%
                  change in carbon line flux density near 8.5 GHz to those obtained using
                  the line flux density given in Table 2.} 
\tablenotetext{c}{Carbon line flux density near 8.5 GHz reduced by 20\% of that given
                  in Table 2.} 
\tablenotetext{d}{Carbon line flux density near 8.5 GHz increased by 20\% of that given
                  in Table 2.} 
\end{deluxetable}

\begin{figure}
\plotone{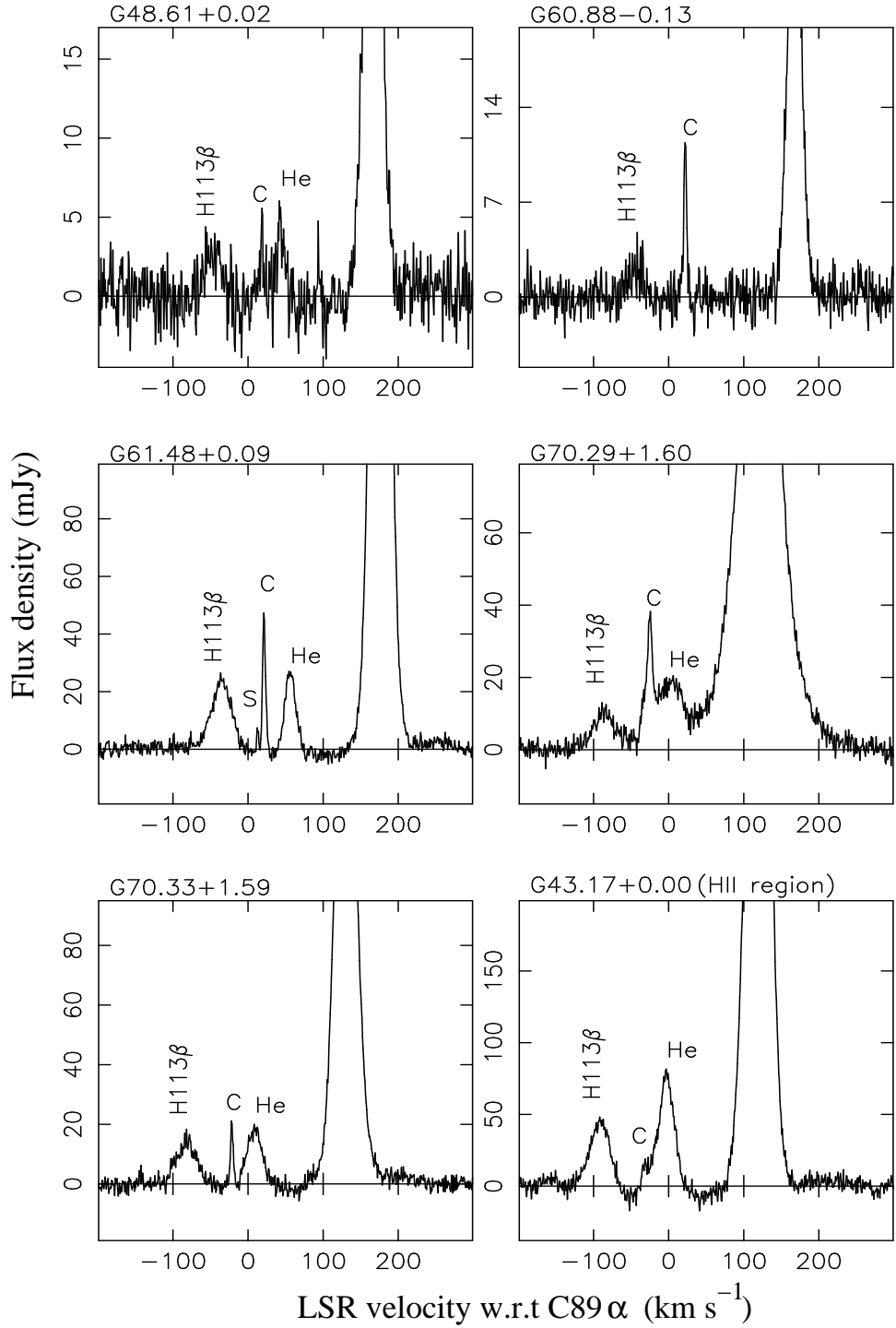}
\caption{(Figure 2 continued)}
\end{figure}

\begin{figure}
\plotone{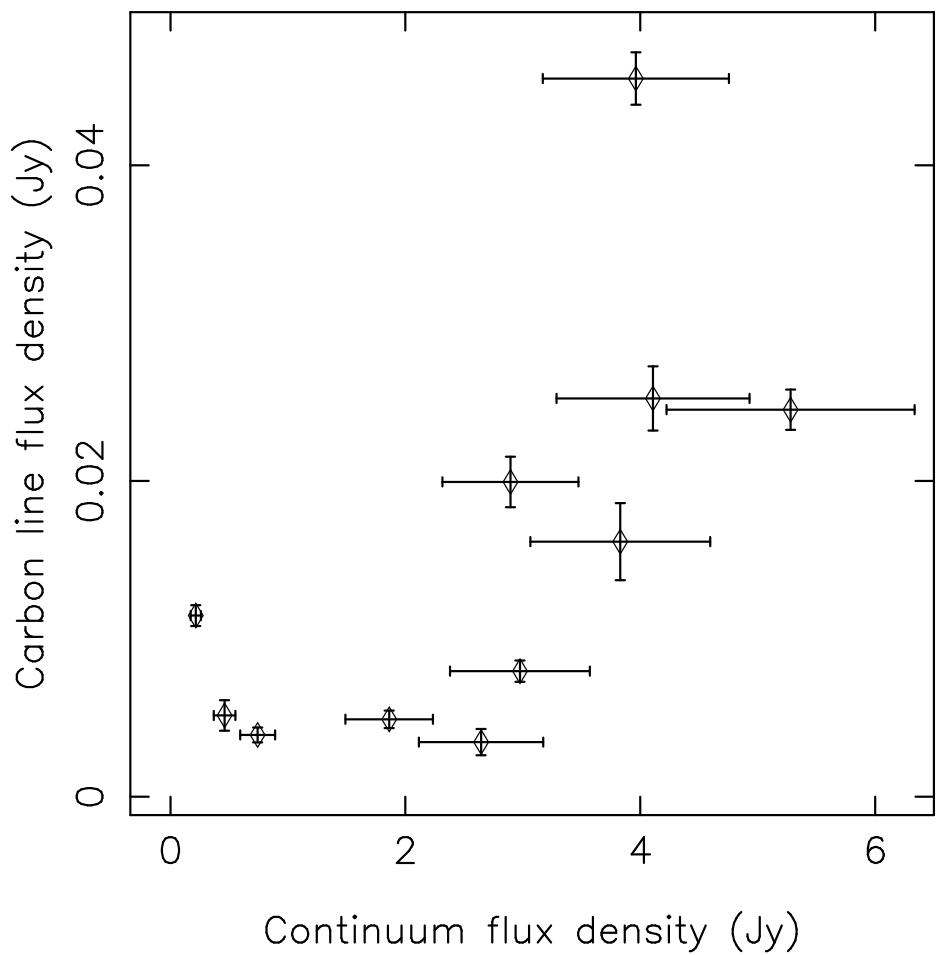}
\caption{
Flux density of carbon RLs observed toward \UCHII s plotted against 
the continuum flux density. The line 
intensity correlates well with the continuum emission indicating the dominance
of stimulated emission for carbon RLs. The correlation 
coefficient is 66\%. The error bar on carbon RL flux density is
the $\pm 1\sigma$ error obtained from the Gaussian modeling of line profiles.
For continuum flux density the error bar is $\pm$ 20\%.
\label{fig2} }
\end{figure}

\begin{figure}
\plotone{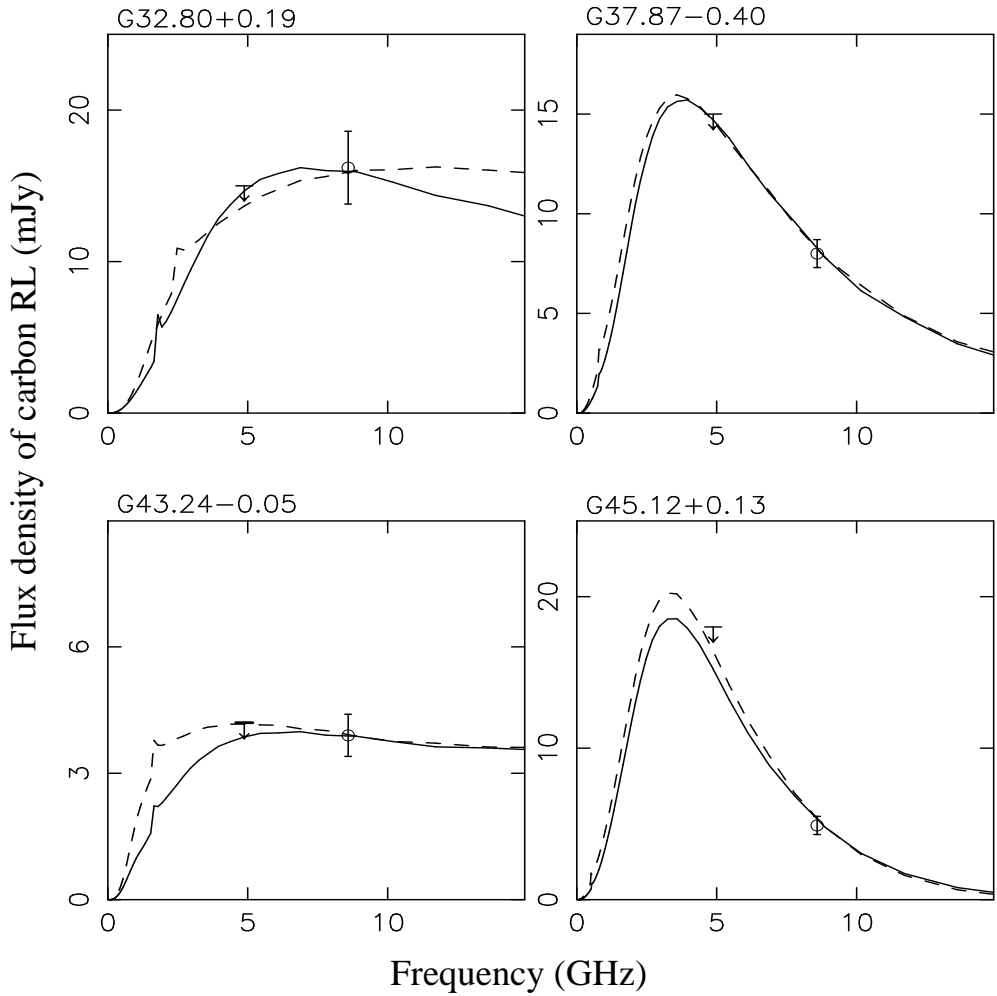}
\caption{Carbon line flux density as a function of frequency from models toward
a subset of \UCHII s in the Arecibo survey. The electron temperatures used for
modeling are 300 K (solid line) and 500 K (dashed line). The electron
density and PDR thickness of the models are given in Table~\ref{tab4}. The curves
are obtained for model parameters that produce the observed carbon line 
flux density near 8.5 GHz 
(marked as circle with $\pm$ 1$\sigma$ error obtained from Gaussian modeling
of line profiles) and are consistent wit the 3$\sigma$ upper limit of carbon
line flux density observed near 4.8 GHz.
\label{fig3} }
\end{figure}

\begin{figure}
\plotone{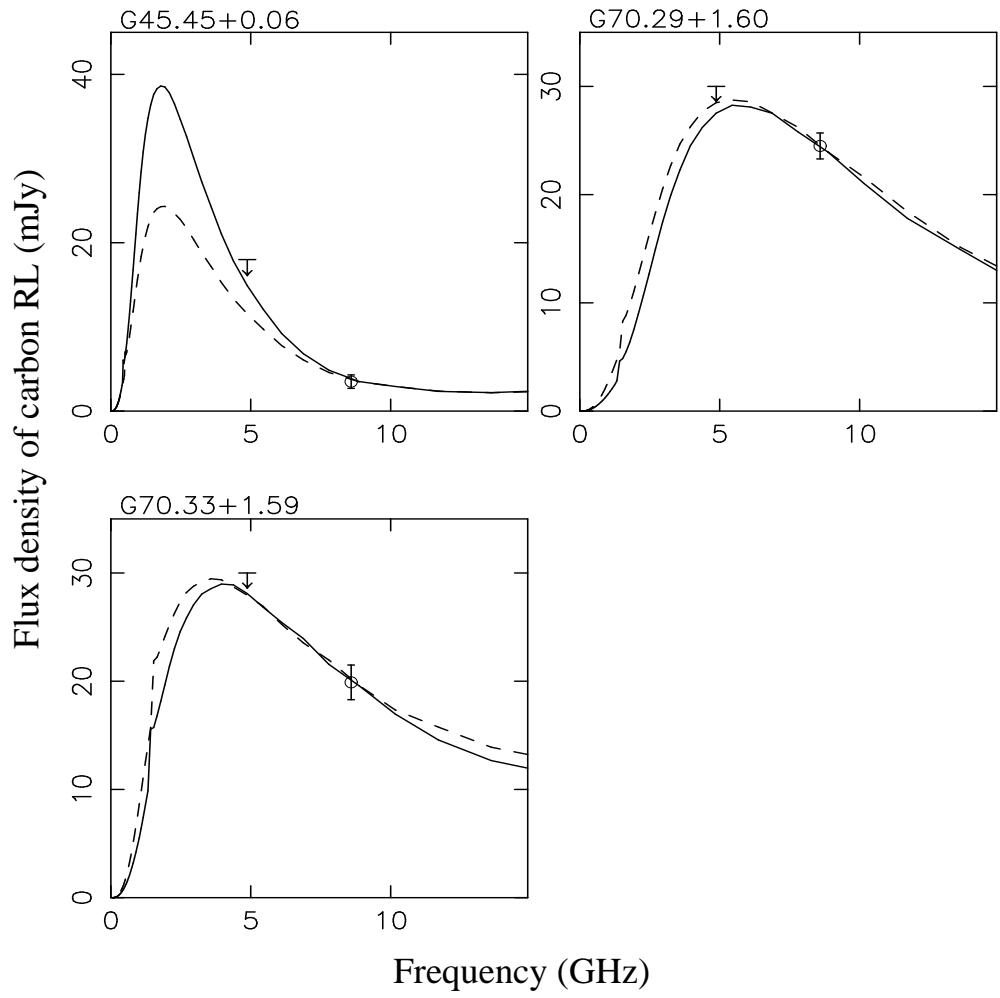}
\caption{(Figure 4 continued)}
\end{figure}

\begin{figure}
\plotone{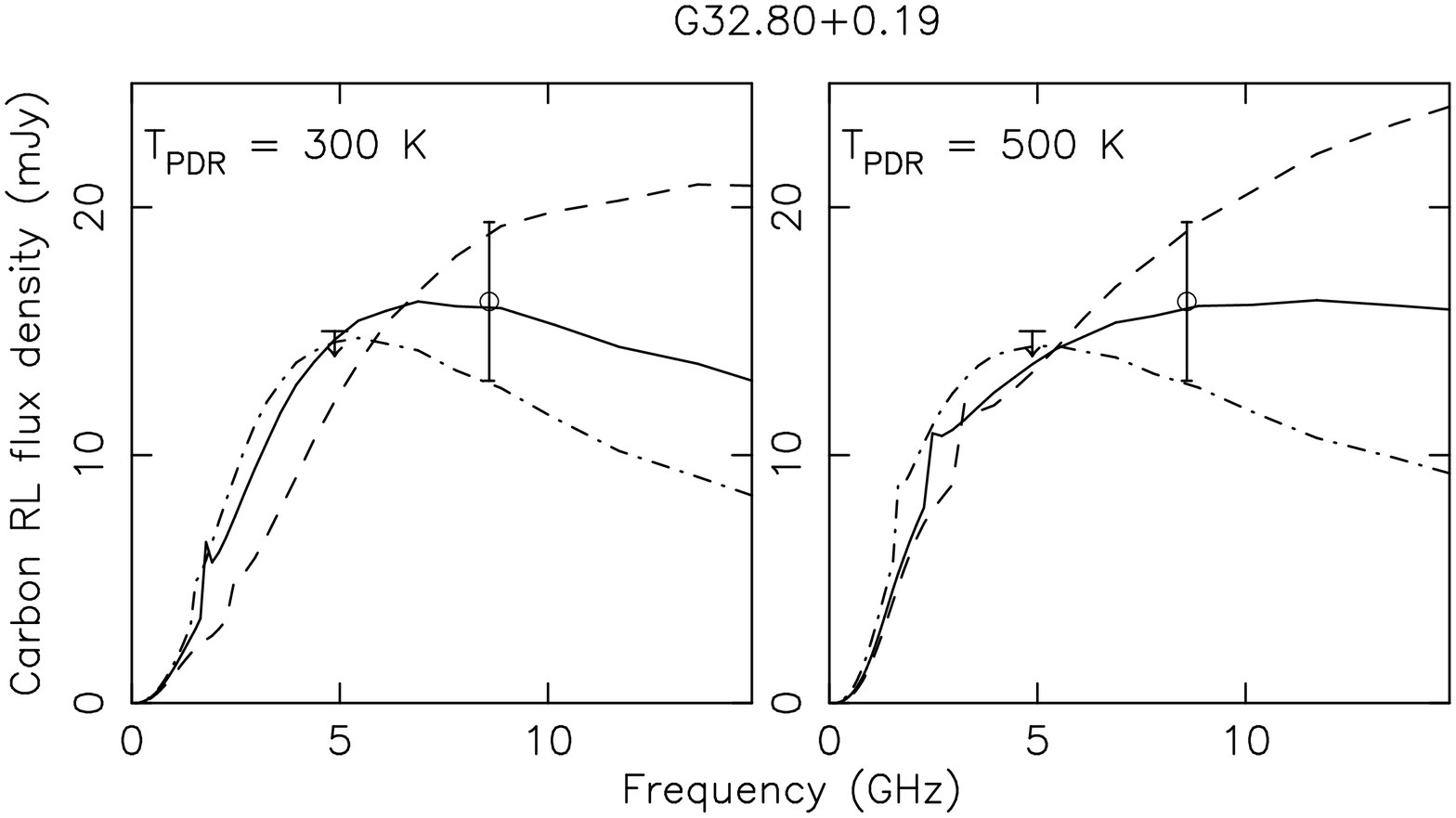}
\caption{
%The plot demonstrates the robustness of the estimated PDR parameters toward G32.80+0.19. 
Carbon line flux density as a function of frequency 
for PDR models toward G32.80+0.19. Model line flux density with gas temperatures 
300 K (left) and 500 K (right) are plotted. The solid line
corresponds to the electron density and PDR thickness given in Table~\ref{tab4} for
G32.80+0.19. Line flux density for model parameters (see Table~\ref{tab5}) that 
are consistent with the upper limit at 4.8 GHz and RL flux densities of 19.4 
and 13.0 mJy at 8.5 GHz are also shown by the `dash' and `dash-dot' lines respectively. 
These RL flux densities are $\pm$ 20 \% of the value given in Table~\ref{tab4}. 
The carbon line flux density near 8.5 GHz is 
marked as circle with $\pm$ 20\% error bar. 
The 3$\sigma$ upper limit of carbon line flux density observed near 4.8 GHz 
is also shown in the plots.  The lower limit on electron density reduces 
only by about a factor of two from that given in Table~\ref{tab4} when the carbon line 
flux density is reduced by 20\% (see \S\ref{sec:rob}). 
\label{fig4} }
\end{figure}

\end{document}